\documentclass[12pt,a4paper,english]{JHEP3}

\usepackage{amsmath,cite,babel,graphicx}
\usepackage[latin1]{inputenc}

\newcommand{\newc}{\newcommand}

\newc{\ptmin}{p_t^{\rm min}}
\newc{\Sinc}{\sigma^{\rm inc}}
\newc{\sigmasoft}{\sigma^{\rm inc}_{\rm soft}}
\newc{\sigmahard}{\sigma^{\rm inc}_{\rm hard}}
\newc{\sigmatot}{\sigma_{\rm tot}}
\newc{\ASinc}{A(b)\cdot\sigma^{\rm inc}}
\newc{\db}{\mathrm{d}^2b \ }
\newc{\dpt}{\mathrm{d}p_t^2 \ }
\newc{\pt}{p_t}
\newc{\vect}[1]{{\bf #1}}
\newc{\HWPP}{\mbox{\textsf{Herwig++}}}
\newc{\GeV}{\text{ GeV}}
\newc{\TeV}{\text{ TeV}}
\newc{\eik}{\chi(\vect{b}, s)}

\newc{\todo}[1]{\textbf{(todo: #1)}\marginpar{\rule{10mm}{3mm}}}

\renewcommand{\exp}[1]{\: e^{#1} \:}

\newc{\HWPPClass}[1]{\mbox{\href{http://projects.hepforge.org/herwig/doxygen/classHerwig_1_1#1.html}{\textsf{#1}}}}
\newc{\HWPPParameter}[2]{\mbox{\href{http://projects.hepforge.org/herwig/doxygen/#1Interfaces.html\##2}{{\bf
#2}}}}

\def\jim{{\sc Jimmy}}

\graphicspath{{./pics/}}

\title{The Underlying Event and the Total Cross Section from Tevatron to the LHC}

\author{Manuel B\"ahr\\ 
  Institut f\"ur Theoretische Physik, Universit\"at Karlsruhe, 
  Germany; and Department of Physics \& Astronomy,
  University College London}

\author{Jonathan M. Butterworth\\ 
  Department of Physics \& Astronomy,
  University College London}

\author{Michael H. Seymour\\
  School of Physics and Astronomy, University of Manchester; and\\
  Physics Department, CERN, CH-1211 Geneva 23, Switzerland}

\abstract{Multiple partonic interactions are widely used to simulate the
hadronic final state in high energy hadronic collisions, and
successfully describe many features of the data. It is important to make
maximum use of the available physical constraints on such models,
particularly given the large extrapolation from current high energy data
to LHC energies. In eikonal models, the rate of multiparton interactions
is coupled to the energy dependence of the total cross section. Using a
Monte Carlo implementation of such a model, we study the connection
between the total cross section, the jet cross section, and the
underlying event. By imposing internal consistency on the model and
comparing to current data we constrain the allowed range of its
parameters. We show that measurements of the total proton-proton
cross-section at the LHC are likely to break this internal consistency,
and thus to require an extension of the model. 
Likely such extensions are that hard scatters probe a denser matter
distribution inside the proton in impact parameter space than soft
scatters, a conclusion also supported by Tevatron data on double-parton
scattering, and/or that the basic parameters of the model are energy
dependent.}

\keywords{Hadronic Colliders, QCD, Jets, Phenomenological Models,
  Underlying Event} 

\preprint{CERN-PH-TH/2008-129\\KA-TP-13-2008\\MCnet/08/03}

\begin{document}

\section{Introduction}

Hadron-hadron collision events at high energies often contain high
transverse energy jets, which in QCD arise from gluon or quark
(generically, parton) scattering followed by QCD radiation and
hadronization. This model is generally taken to be realistic above
some minimum transverse momentum scale, $\ptmin$. The contribution of
these events to the total cross section rises with hadron-hadron
centre-of-mass energy, $s$, since the minimum value of the $x$ probed
is given by $4(\ptmin)^2/s$, and the parton densities rise strongly
for $x < 10^{-2}$ or so~\cite{Chekanov:2001qu,Adloff:2000qk}. 

One reason that this rising contribution to the cross section is of
interest is that while perturbative QCD cannot predict total hadronic
cross sections (since in many events no hard perturbative scale is
present), rising hadronic cross sections are a feature also seen in
successful non-perturbative approaches~\cite{dl,Donnachie:2004pi}, the
behaviour of which must presumably emerge from the QCD Lagrangian in
some manner. Thus by comparing the behaviour of the hard contribution
to the cross section to the behaviour expected from fits to the total
cross section, it may be possible to learn something about the
connection between these approaches and about hadronic cross sections
in general.

The connection between the hard partonic cross section and the total
cross section is not one-to-one, however. There are certainly hadronic
scatters in which no hard jets are produced, and some non-perturbative
scattering process must be added to the perturbative jet contribution
to model the total cross section. In addition, at the high parton densities
probed at recent, current and future colliders, simple assumptions
lead to the conclusion
that the probability of multiple partonic scatters in a single hadron-hadron collision is significant. In
fact, Fig.~\ref{fig:sigmainc} shows that for $\ptmin$ values below
about 5~GeV, the total ``hard'' cross section calculated assuming one
parton-parton scatter per proton-proton collision {\it exceeds the
total cross section} as extrapolated using the non-perturbative fits,
at LHC energies. This
strongly implies that the average number of partonic scatters in an
inelastic collision must be greater than one.

\FIGURE[htb]{%
  \includegraphics[scale=0.4]{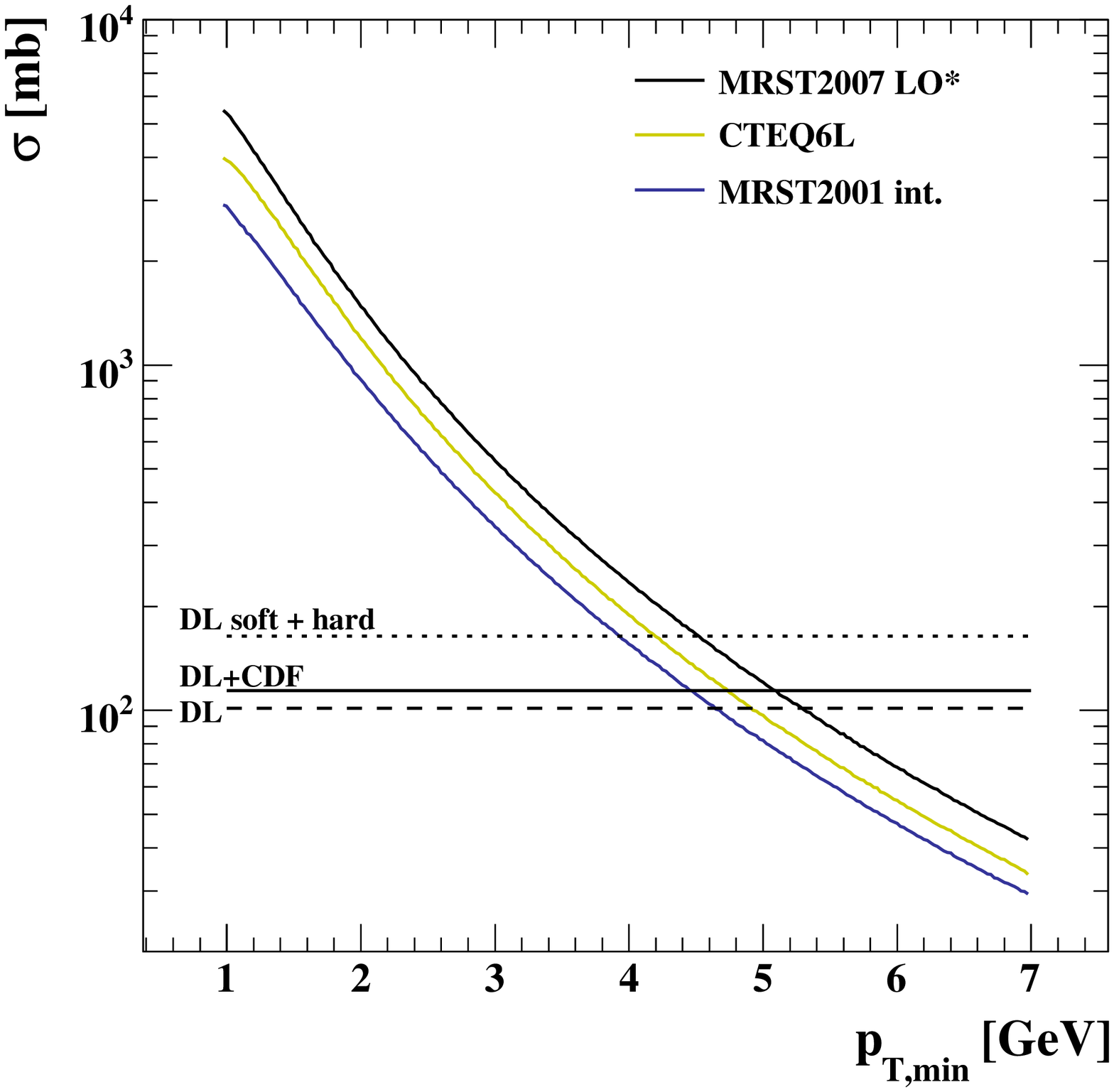}
  \caption{ The inclusive hard cross section for three different proton
    PDFs, compared to various extrapolations of the non-perturbative
    fits to the total $pp$ cross section at 14~TeV centre-of-mass
    energy.}
  \label{fig:sigmainc}
} 

Introducing the possibility of such multiparton interactions also
seems to be required in order to describe the hadronic final state\cite{Sjostrand:1987su,Affolder:2001xt,Chekanov:2007et}. In
general, softer additional scatters occurring in a high-$\pt$\linebreak event
manifest themselves as additional particles and energy-flow, the
so-called ``underlying event''.

In this paper we examine the predictions of the model that was
discussed, for example in~\cite{Durand:1988ax}, and
implemented in~\cite{Butterworth:1996zw,ivan,Bahr:2008dy,Bahr:2008pv}
including\linebreak the possibility of soft scatters. We explore the
consistency constraints that would be imposed by comparing a given value
of the total cross section to
the predicted jet cross section, and attempt to identify allowed regions
of parameter space within which the model must lie if it is to be
consistent with the measured cross section
at the LHC.  We also discuss ways in which energy dependencies in the
parameters could arise, and their impact upon these constraints.  The
studies are all carried out using the new implementation in
\HWPP~\cite{Bahr:2008dy,Bahr:2008pv}; however, they are also relevant to
the fortran implementation \jim~\cite{Butterworth:1996zw}, if the same
hard cross section is used.

\section{Total and elastic cross section parameterizations}\label{sec:DL}

Throughout this paper we will exploit the connection that can be
established between the eikonal model of Refs.~\cite{Butterworth:1996zw,
ivan, Bahr:2008dy} and the total cross section. To give a reasonable
range of expectations for the latter, we use
the successful parameterization of Donnachie and Landshoff
\cite{dl, Donnachie:2004pi}. We will use three different variations;

\begin{enumerate}
  \item The \emph{standard} parameterization from \cite{dl} with the
    following behaviour at high energies:
    \begin{align}
      \sigmatot \sim 21.7 \ \mathrm{mb} \cdot
      \left(\frac{s}{\GeV^2}\right)^{0.0808} \ \rightarrow \
      \sigmatot(14\TeV) = 101.5 \ \mathrm{mb} \, .
    \end{align}

  \item Using the same energy dependence but normalizing it to the
    measurement \cite{sigma_tot_CDF} by CDF:
    \begin{align}
      \sigmatot \sim 24.36 \ \mathrm{mb} \cdot
      \left(\frac{s}{\GeV^2}\right)^{0.0808} \ \rightarrow \
      \sigmatot(14\TeV) = 114.0 \ \mathrm{mb} \, .
    \end{align}

  \item Using the most recent fit \cite{Donnachie:2004pi}, which takes
    the contributions from both hard and soft Pomerons into account:
    \begin{align}
      \begin{split}
	&\sigmatot \sim 
	24.22 \ \mathrm{mb} \cdot \left(\frac{s}{\GeV^2}\right)^{0.0667} +
	0.0139 \ \mathrm{mb} \cdot \left(\frac{s}{\GeV^2}\right)^{0.452} \\
	\rightarrow \
	&\sigmatot(14\TeV) = 164.4 \ \mathrm{mb} \, .
      \end{split}
    \end{align}
\end{enumerate}
Other parameterizations and models for the total cross section
exist~\cite{Khoze:2000wk,Gotsman:2007ac}, but their predictions for the
total cross section at 14~TeV generally lie within the range covered by
these three\footnote{The most recent models of \cite{Ryskin:2007qx,
Gotsman:2008tr} predict $\sigmatot(14\TeV) \simeq 90$ mb, which is 10 \%
below the smallest expectation we use. Since the difference this
introduces is similar to the one between our first and second
parameterization it can easily be estimated by the reader.}. As will be
seen, the range is wide, and early measurements of the total cross
section at the LHC can be expected to have a big
impact~\cite{Deile:2006tt}.

We will also find it useful to compare our model with the elastic slope
parameter, $B$, defined in terms of the differential elastic scattering
cross section, $\text{d}\sigma/\text{d}t$, as
\begin{equation}
  B = B(s,t=0) = \left[ \frac{\text{d}}{\text{d}t} \left( \ln
  \frac{\text{d}\sigma}{\text{d}t}\right) \right]_{t=0}.
\end{equation}
In the Donnachie-Landshoff parameterization, this is given by:
\begin{equation}
  B = 2\alpha'\ln\frac{s}{s_0} + B_0
\end{equation}
with $\alpha'=0.25\GeV^{-2}$.  Together with the CDF data\cite{Abe:1993xx},
this implies
\begin{equation}
  B = \left(\ln\frac{\sqrt{s}}{1800\GeV}+(17\pm0.25)\right)\GeV^{-2}
  = \left(\ln\frac{\sqrt{s}}{14\TeV}+(19\pm0.25)\right)\GeV^{-2}.
\end{equation}
The most recent fit\cite{Donnachie:2004pi} has the same value for
$\alpha'$ and hence $B$, while those of
\cite{Khoze:2000wk,Gotsman:2007ac} are a little higher:
20--22$\GeV^{-2}$.  We therefore use the CDF value for the Tevatron
energy and the range 19--22$\GeV^{-2}$ to represent the range of
possible measurements from the LHC.

\section{Eikonal model}

The scattering amplitude $\mathcal{A}(s,t)$ can be expressed as the
Fourier transform of the elastic scattering amplitude $a(\vect{b},s)$ in
impact parameter space as
\begin{equation}
  \label{eq:fourier}
  \mathcal{A}(s,t) = 4 s \int \db  \ a(\vect{b}, s)
  \exp{i \vect{q} \cdot \vect{b}}  \, ,
\end{equation}
where $\vect{q}$ is the transverse momentum transfer vector, with, in
the high energy limit, $\vect{q}^2=-t$.  In this limit, $a(\vect{b},s)$
can be assumed to be purely imaginary and therefore be expressed in
terms of a real eikonal function $\eik$, as
\begin{equation}
  \label{eq:eikonal}
  a(\vect{b},s) = \frac{1}{2 i} \left[ \exp{-\eik} - 1 \right] \, .
\end{equation}
Using (\ref{eq:fourier}) and (\ref{eq:eikonal}) the total cross
section for $pp\to X$ can be expressed as
\begin{equation}
  \label{eq:sigma_tot}
  \begin{split}
    \sigmatot = \ & \frac{1}{s} \Im\left\{ \mathcal{A}(s, t=0) \right\}
    \\
    = \ & 2 \int \db \ \left[ 1 - \exp{-\eik} \right] \, .
  \end{split}
\end{equation}
The elastic cross section is then
\begin{equation}
  \label{eq:sigma_ela}
  \begin{split}
    \sigma_{\rm el} = \ & 4 \int \db \ \left| a(\vect{b}, s) \right|^2 \\
    = \ &  \int \db \ \left| 1 - \exp{-\eik} \right|^2 \, .
  \end{split}
\end{equation}
The inelastic cross section thereby reads
\begin{equation}
  \label{eq:sigma_inel}
  \begin{split}
    \sigma_{\rm inel} = \ & \sigmatot - \sigma_{\rm el} \\
    = \ & \int \db \ \left[ 1 - \exp{-2\eik} \right] \, .
  \end{split}
\end{equation}
The elastic slope parameter at zero momentum transfer
is also calculable within this framework and yields \cite{Block:1984ru}
\begin{equation}
  B = \frac{1}{\sigmatot} \int \db b^2  \ \left[ 1 - \exp{-\eik} \right] \, .
\end{equation}

\subsection{Multi-parton scattering model}

The preceding expressions are completely general and model-independent,
but we now introduce an explicit model\cite{Durand:1988ax,
Butterworth:1996zw, ivan, Bahr:2008dy} to predict the form of the
eikonal function, $\eik$.  First we assume that it can be decomposed
into the sum of independent soft and hard parts,
\begin{equation}
  \label{eq:eik_tot}
  \chi_{\rm tot}(\vect b,s) = \chi_{\rm QCD}(\vect b,s) + \chi_{\rm soft}(\vect b,s) \, ,
\end{equation}
and start by considering the hard part.  We consider a model in which
partons are distributed across the face of the colliding hadrons with a
spatial distribution that is independent of their longitudinal
momentum.  We assume that pairs of partons in the colliding hadrons
scatter with independent probabilities, leading to the
distribution of number of scatters at fixed impact parameter obeying
Poisson statistics.  We further assume that any hadron-hadron collision
in which there is an elastic parton-parton collision above some cutoff
$\ptmin$ will lead to an inelastic hadronic final state.  Finally, we
require that the inclusive cross section for hadron-hadron collisions to
produce partons above $\ptmin$ be equal to the inclusive partonic cross
section folded with standard inclusive parton distribution functions, as
given by the factorization theorem.  The result of this model is that
the inelastic cross section is given by an expression identical to
Eq.~\ref{eq:sigma_inel}, but with
$\chi$ replaced by
\begin{equation}
  \label{eq:eik_hard}
  \chi_{\rm QCD}(\vect{b},s) = \frac{1}{2}\, A(\vect{b}) \
  \sigmahard(s;\ptmin) \, ,
\end{equation}
where $A(\vect{b})$ describes the overlap distribution of the partons in
impact parameter space and $\sigmahard$ denotes the inclusive cross
section above a transverse momentum cutoff $\pt > \ptmin$, given by the
perturbative result
\begin{equation}
  \sigmahard(s;\ptmin) = \sum_{ij}\int \mathrm{d}x_1 \, \mathrm{d}x_2 \;
  f_i(x_1) \, f_j(x_2) \, \int_{\ptmin} \mathrm{d}\pt \,
  \frac{\mathrm{d}\hat\sigma_{ij}(x_ix_js)}{\mathrm{d}\pt},
\end{equation}
where $\mathrm{d}\hat\sigma_{ij}(\hat s)/\mathrm{d}\pt$ denotes the
inclusive cross section for partons of types $i$ and $j$ and
invariant-mass-squared $\hat s$ to produce jets of a given $\pt$.

We assume that the soft eikonal function has the same form,
\begin{equation}
  \label{eq:eik_soft}
  \chi_{\rm soft}(\vect{b},s) = \frac{1}{2}\, A_{\rm soft}(\vect{b}) \ \sigmasoft \, ,
\end{equation}
where $\sigmasoft$ is the purely non-perturbative cross section below
$\ptmin$, which is a free parameter of the model. That is, we assume
that soft scatters are the result of partonic interactions that are
local in impact parameter.

The elastic slope parameter discussed above relates to bulk interactions
of the proton. Thus it can be taken as directly constraining the matter
distribution ``seen'' by soft scatters. Higher $\pt$ scatters might be
expected to see a different matter distribution, for example they might
probe denser ``hot spots'' within the proton. However, at present we
take the simplest assumption for the perturbative part of the eikonal
function, i.e. that the semi-hard scatters ``see'' the same matter
distribution as the soft ones and therefore take $A(\vect b) \equiv 
A_{\rm soft}(\vect b)$. This is clearly a simplifying assumption, but a
reasonable one until proven otherwise.

It is worth mentioning that according to the definition in
Eq.~\ref{eq:sigma_inel}, the inelastic cross section contains
\emph{all\/} inelastic processes, including diffractive dissociation.
This is consistent with the calculation of the inclusive hard cross
section, which is calculated from the conventional parton distribution
functions, which describe the \emph{inclusive\/} distribution of partons
in a hadron whatever their source, i.e.\ whether the proton remains
intact or not.

\subsection{Overlap parameterization}

The only remaining freedom in the eikonal model is the functional form
of the overlap function $A(|\vect b|=b)$. $A(b)$ is the convolution of
the individual spatial parton distributions of the colliding hadrons,
\begin{align}
  A(b) = \int d^2\vect{b}^{\prime} \ G_{h_1}(|\vect{b}^\prime|) \
  G_{h_2}(|\vect{b} - \vect{b}^\prime|)  \, .
\end{align}
In Refs.~\cite{Butterworth:1996zw, ivan, Bahr:2008dy}, $G(\vect b)$ is
taken to be proportional to the electromagnetic form factor,
\begin{equation}
  G_{\bar p}(\vect{b}) = G_p(\vect{b}) = \int \frac{d^2\vect{k}}{(2\pi)^2} \
  \frac{\exp{i \vect{k} \cdot \vect{b}}}{(1+\vect{k}^2/\mu^2)^2} \ .
\end{equation}

\FIGURE[htb]{\label{fig:overlap} 
  \includegraphics[scale=0.37]{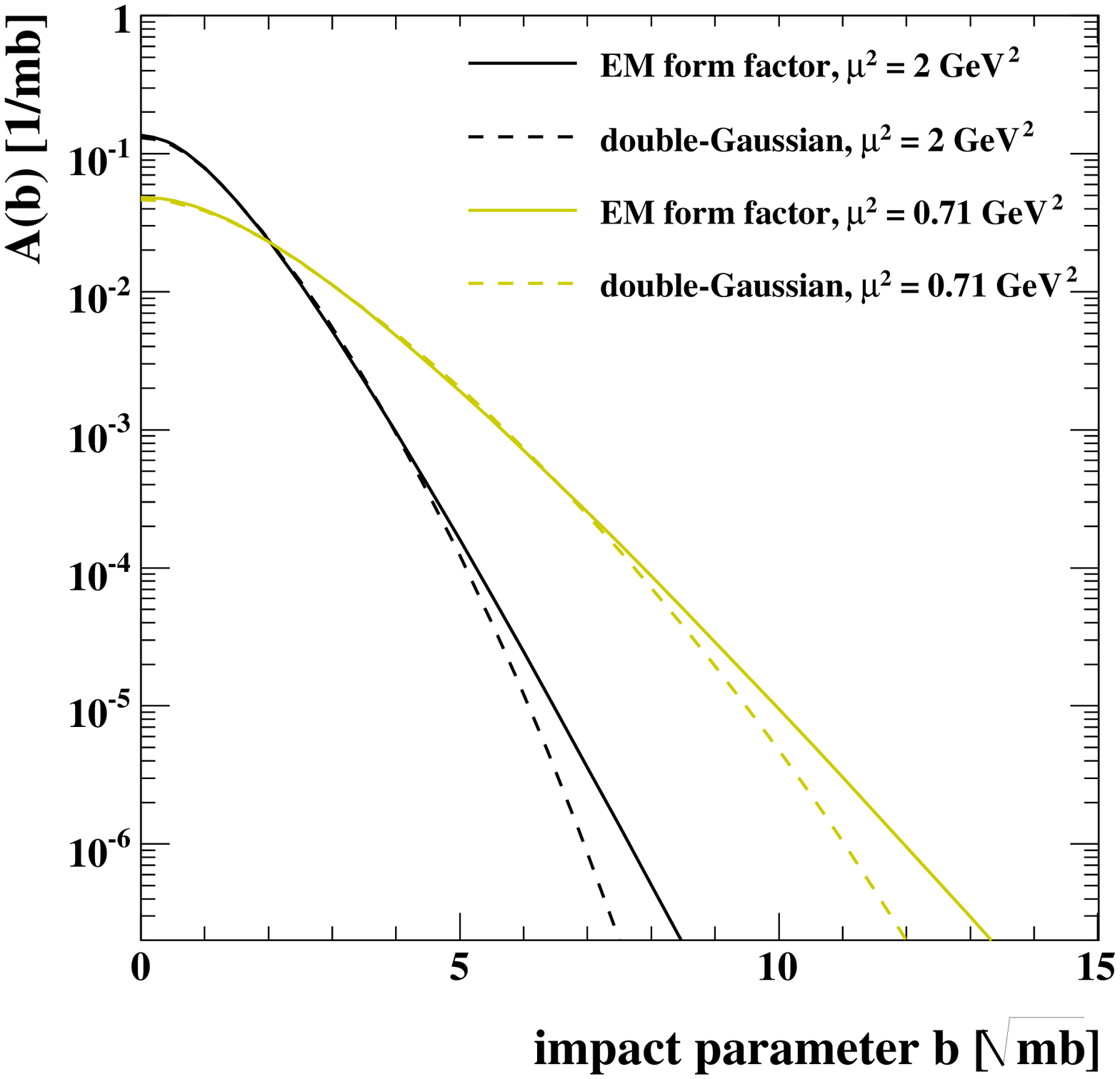}
  \vspace*{-0.2cm}
  \caption{$A(b)$ for the two 
    parameterizations.}  }

\sloppypar
\noindent
$\mu$ is the only parameter and has the dimensions of an inverse radius.  In
$ep$ scattering its value was measured to be $\mu^2=0.71
\GeV^2$. This is a loose constraint, since the distribution of
partons may not necessarily coincide with the distribution of
electromagnetic charge. Actually, using the results from the
previous section, the CDF data on the total cross section
($\sigmatot=81.8\pm2.3$~mb \cite{sigma_tot_CDF}) and the elastic slope
($B = 16.98 \pm 0.25 \GeV^{-2}$ \cite{Abe:1993xx}) one can solve for the
total inclusive cross section and for $\mu^2$, yielding $\mu^2 = 0.56 \pm
0.01 \GeV^2$.

In order to investigate the dependence on the assumed shape of the
matter distribution, we have compared our default results with those
obtained with a double-Gaussian distribution, as chosen in
Refs.~\cite{Sjostrand:2004pf,Sjostrand:2006za,Sjostrand:2007gs},\\[0.3cm]
\begin{equation}
  G(b) = \frac{1-\beta}{\pi r^2} \cdot \exp{-\frac{b^2}{r^2}} +
  \frac{\beta}{\pi (k\cdot r)^2} \cdot \exp{-\frac{b^2}{(k\cdot r)^2}}
  \, .
\end{equation}

\noindent
Here $\beta, k$ and $r$ are all free parameters, but we choose to fix
$\beta$ and $k$ at values that are reasonably generic, but also close to
the tuned values used in
\cite{Sjostrand:2004pf,Sjostrand:2006za,Sjostrand:2007gs}, with the
relative strengths given by $\beta = 0.5$ and the relative widths by
$k=2$, and view $r$ as the only free parameter.  The parameters $\mu^2$
and $r$ in the two models are arbitrary and should ultimately be fit to
data.  However, in order to have a like-for-like comparison, we choose
to relate them in such a way that the rms value of $G(b)$ is identical.
That is, we describe the double-Gaussian also as being a function of
$\mu^2$, with $r$ set via $b_{rms}$. We illustrate the shapes of the two
resulting overlap functions for two different values of $\mu^2$ in
Fig.~\ref{fig:overlap}.

We find that for small values of $\mu^2$ ($\sim1 \GeV^2$) the results of
the two models are extremely similar, differing at most by $\pm 2$~\%.
For large values ($\sim3 \GeV^2$) they differ more, the double Gaussian
distribution giving a larger de-eikonalized cross section (see next
section) by between 30 \% with the standard Donnachie--Landshoff total
cross section prediction at the LHC and 150 \% with the soft+hard Pomeron
prediction. While these lead to somewhat different predictions, in our
final results they effectively correspond to a distortion of the $\mu^2$
axis. The effect on our final plots, Figure~\ref{fig:constraintslhc}, is
small, since our consistency requirement is mainly active at small~$\mu^2$.

\subsection{Connection to the total cross section}\label{sec:connection}

\FIGURE[t]{
  \centerline{
    \includegraphics[width=0.47\textwidth]{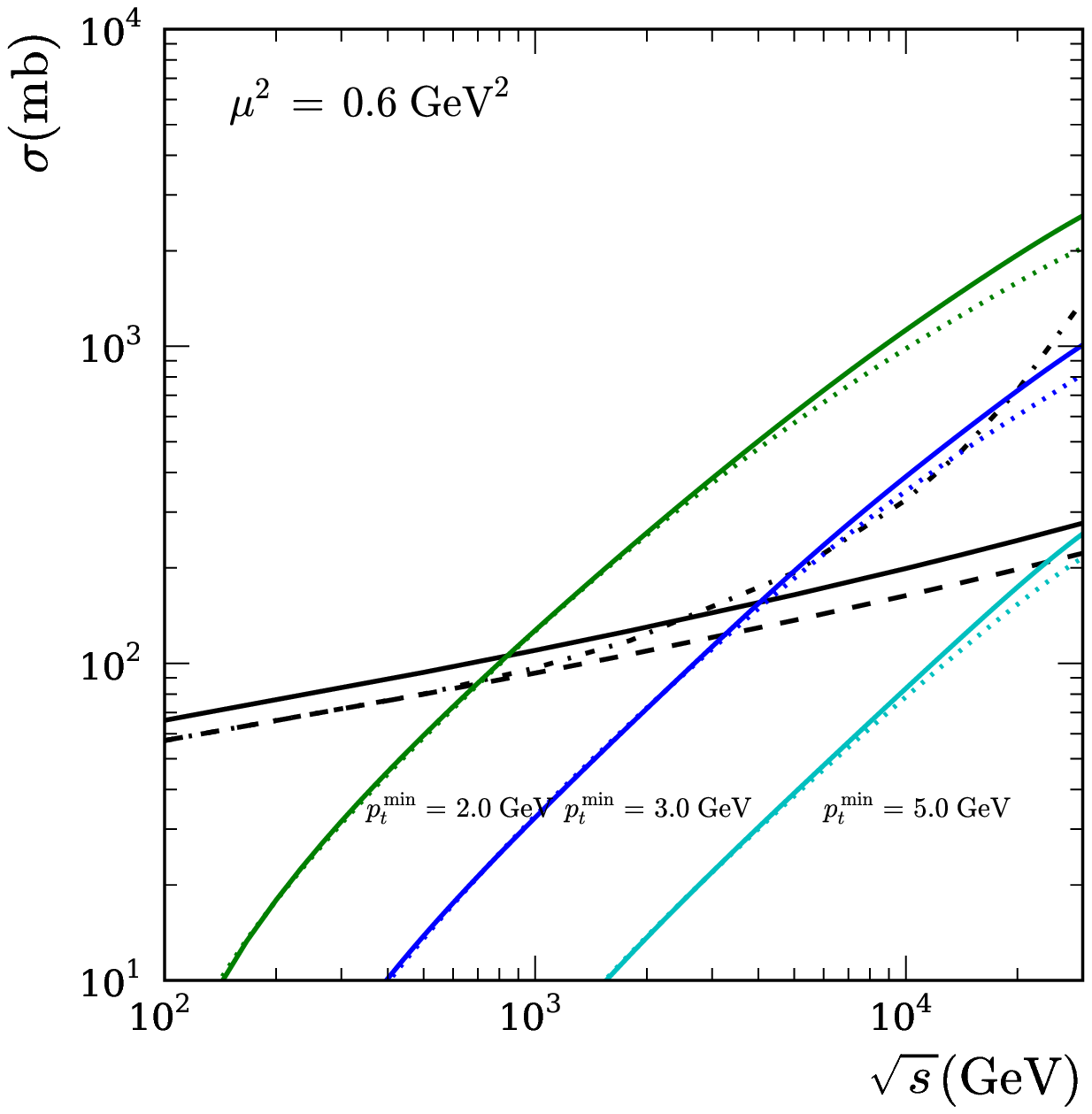}
    \includegraphics[width=0.47\textwidth]{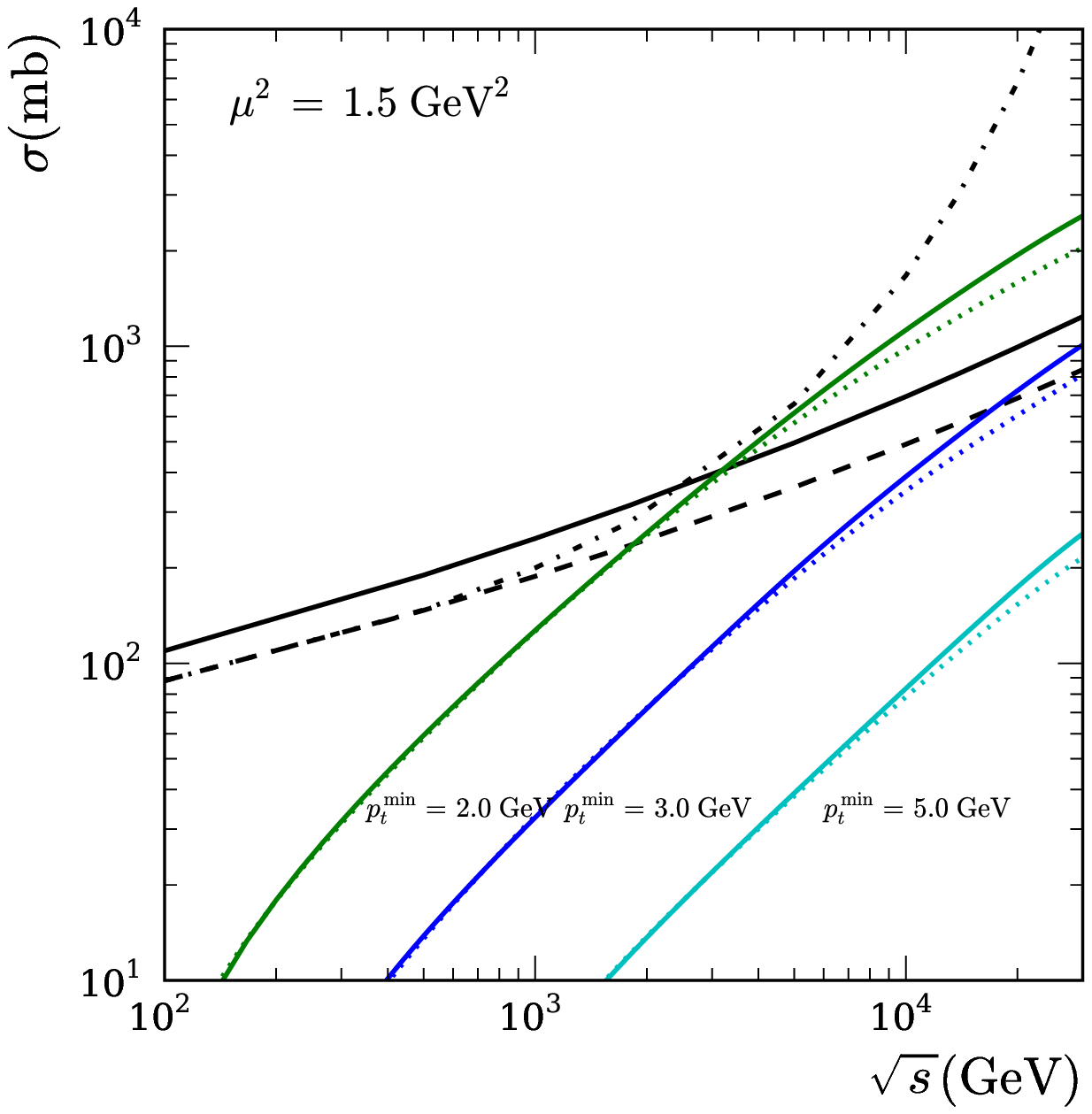}
  }\\
  \centerline{
    \includegraphics[width=0.47\textwidth]{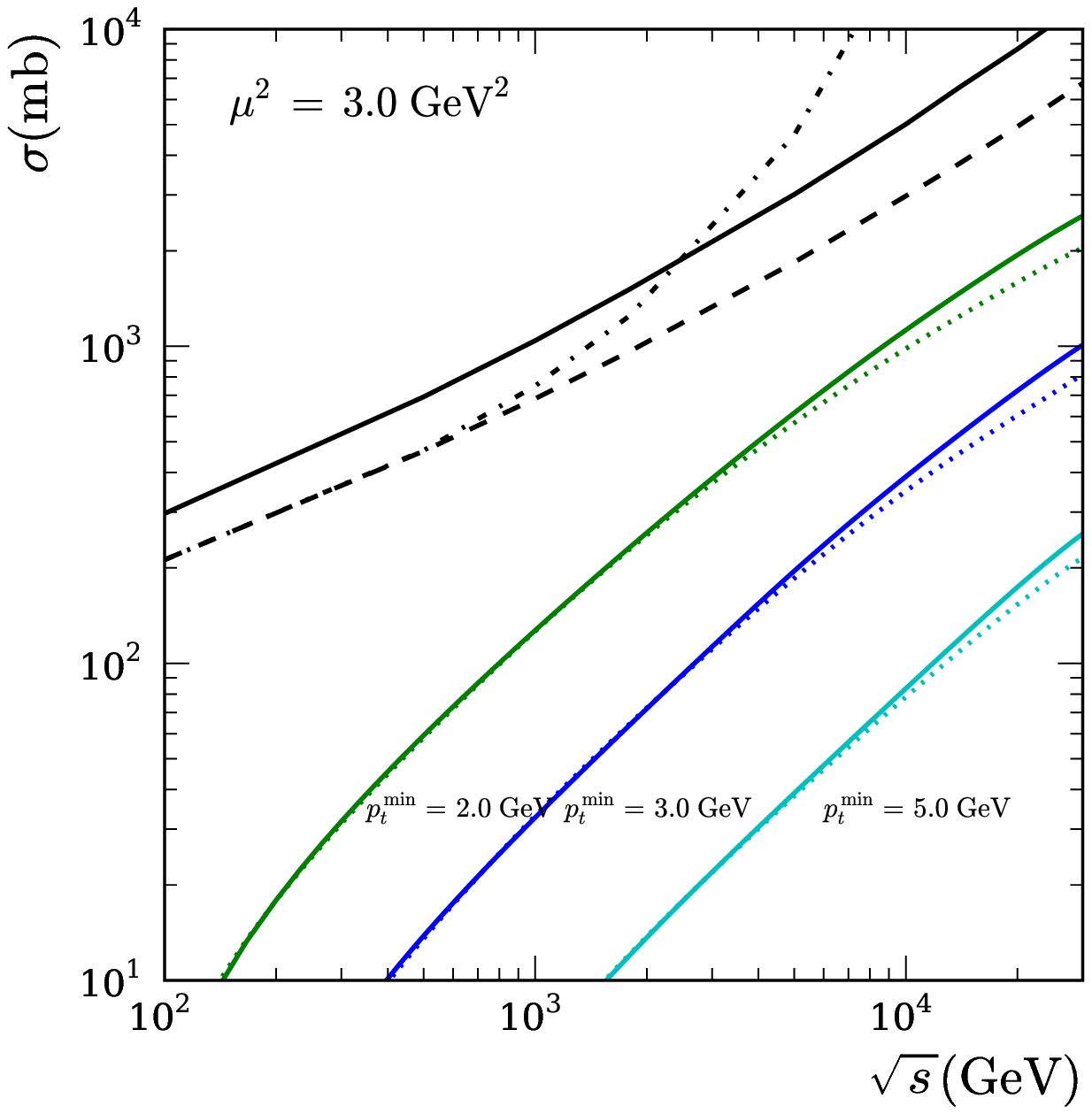}
    \includegraphics[width=0.47\textwidth]{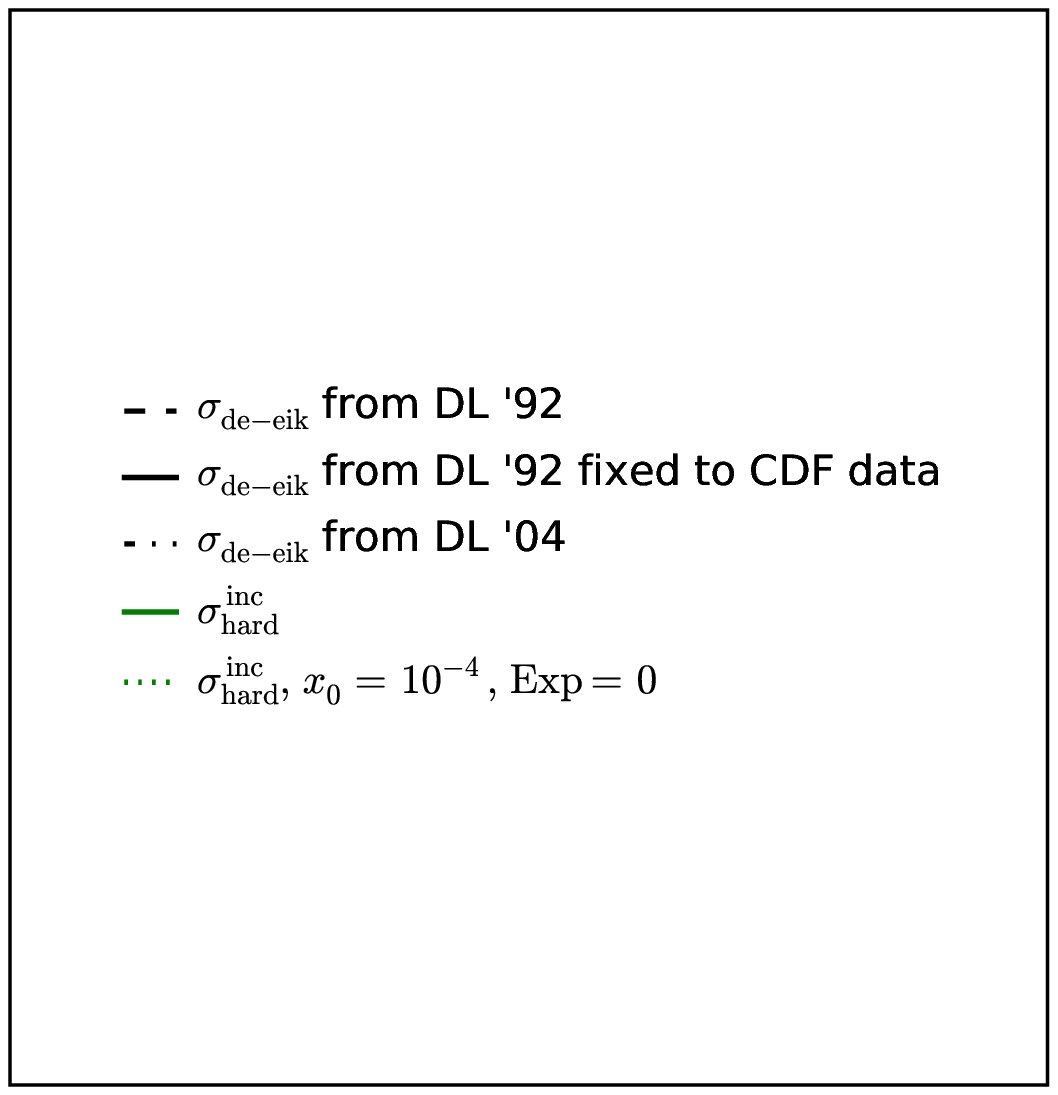}
  }
  \caption{ Cross sections in millibarn as a function of the
    centre-of-mass energy in GeV. The three different plots vary the
    value of $\mu^2$ from $0.6$ to $3 \GeV^2$. The black curves
    show de-eikonalized total cross sections. We use the total cross
    section parameterization of Ref.~\cite{dl} for the dashed curves. The
    solid curves use the same exponent, but the normalization is
    rescaled to fit the total cross section measurement of CDF. The
    dotted curve uses the parameterization of
    Ref.~\cite{Donnachie:2004pi}. The coloured solid curves show
    $\sigmahard$ for different values of $\ptmin$. The coloured
    dash-dotted curves incorporate the simple small-$x$ deviations
    discussed in Sect.~\ref{sec:smallx}}
  \label{fig:deeik}
} 

For a given point in the parameter space ($\ptmin, \mu^2$) of our model,
we are able to calculate $\chi_{\rm QCD}$ using Eq.~\ref{eq:eik_hard}. The
remaining uncertainty is in $\sigmahard(s;\ptmin)$, which depends on the
PDF choice, the treatment of $\alpha_s$ etc.  If we now choose a value
for the non-perturbative cross section below $\ptmin$,
$\sigmasoft$, we have the full eikonal function at hand and can
calculate the total cross section from Eq.~\ref{eq:sigma_tot}.

We will, however, turn this argument around and use the value of
the total cross section as input to fix
the additional parameter in our non-perturbative part of the eikonal
function ($\sigmasoft$). For energies at which there are not yet
measurements, we use the parameterizations of \cite{dl, Donnachie:2004pi}
to give an indication of the likely range of the total cross section.
We will extract the sum $\sigmahard + \sigmasoft \equiv \sigma_{\rm
  de-eik}$ from Eq.~\ref{eq:sigma_tot} and call this cross section the
\emph{de-eikonalized\/} cross section.  That is, the de-eikonalized
cross section is given by the solution to
\begin{equation}
    \sigmatot = 2 \int \db \ \left[ 1 - \exp{-\frac{1}{2} A(\vect{b})
    \sigma_{\rm de-eik}} \right] \, ,
\end{equation}
for a given value of $\mu^2$ and a given value of the total cross
section, $\sigmatot$.  Clearly, $\sigma_{\rm de-eik}$ is a function
only of these two quantities.  Since $\sigmahard$ is $\ptmin$-dependent,
this implies that the value of $\sigmasoft$ we extract by this procedure
is also $\ptmin$-dependent ($\ptmin$ is a \emph{matching\/} scale
between the two sub-process cross sections and the sum of the two is
independent of it).

In Fig.~\ref{fig:deeik} we plot the
de-eikonalized cross sections for the three different total cross
section extrapolations as a function of the centre-of-mass
energy. Furthermore we show the value of $\sigmahard$ using different
cutoffs. $\sigmasoft$ is now given by the difference of these
curves. If we interpret $\sigmasoft$ as a physical cross section (the
inclusive cross section for two partons to undergo a non-perturbative
soft scattering), it cannot be negative. Thus the implication is that
whenever the inclusive hard cross
section is larger than the de-eikonalized one, the model is inconsistent.
We will investigate this
behaviour in more detail in Sect.~\ref{sec:Consistency}.

From the plots in Fig.~\ref{fig:deeik} the values for $\sigmasoft$ can
in principle be read off. However, due to the logarithmic scale it is
not easy to see what is implied for the energy dependence of the soft
cross section. Therefore, for selected points in parameter space,
$\sigmasoft$ is shown separately in Fig.~\ref{fig:sigma_soft}. Note that
where the inclusive hard cross section line for $\ptmin = 3.0$~GeV
crosses and re-crosses the total cross-section extrapolation in the top
left plot of Fig.~\ref{fig:deeik}, the soft cross section in the top
left plot of Fig.~\ref{fig:sigma_soft} first becomes negative and then
positive again.  The dependence of $\sigmasoft$ on the centre-of-mass
energy reveals two main points: First, it is noticeable that one
observes a more-or-less constant behaviour with increasing energy only
in a limited range of our parameter space. This behaviour is mainly
present for lower values of $\mu^2$. Second, for the most extreme total
cross section prediction, $\sigmahard$ is {\it never} sufficient to
explain the strong rise with energy. There, essentially all parameter
choices require a strongly rising {\it soft} cross section, in addition
to the expected strong rise in the hard cross section. This is, at the
very least, counter-intuitive, and one might conclude that, within our
model, the rise of the cross section in parametrization (2.3) is too
extreme.

\FIGURE[t]{
  \centerline{
    \includegraphics[width=0.47\textwidth]{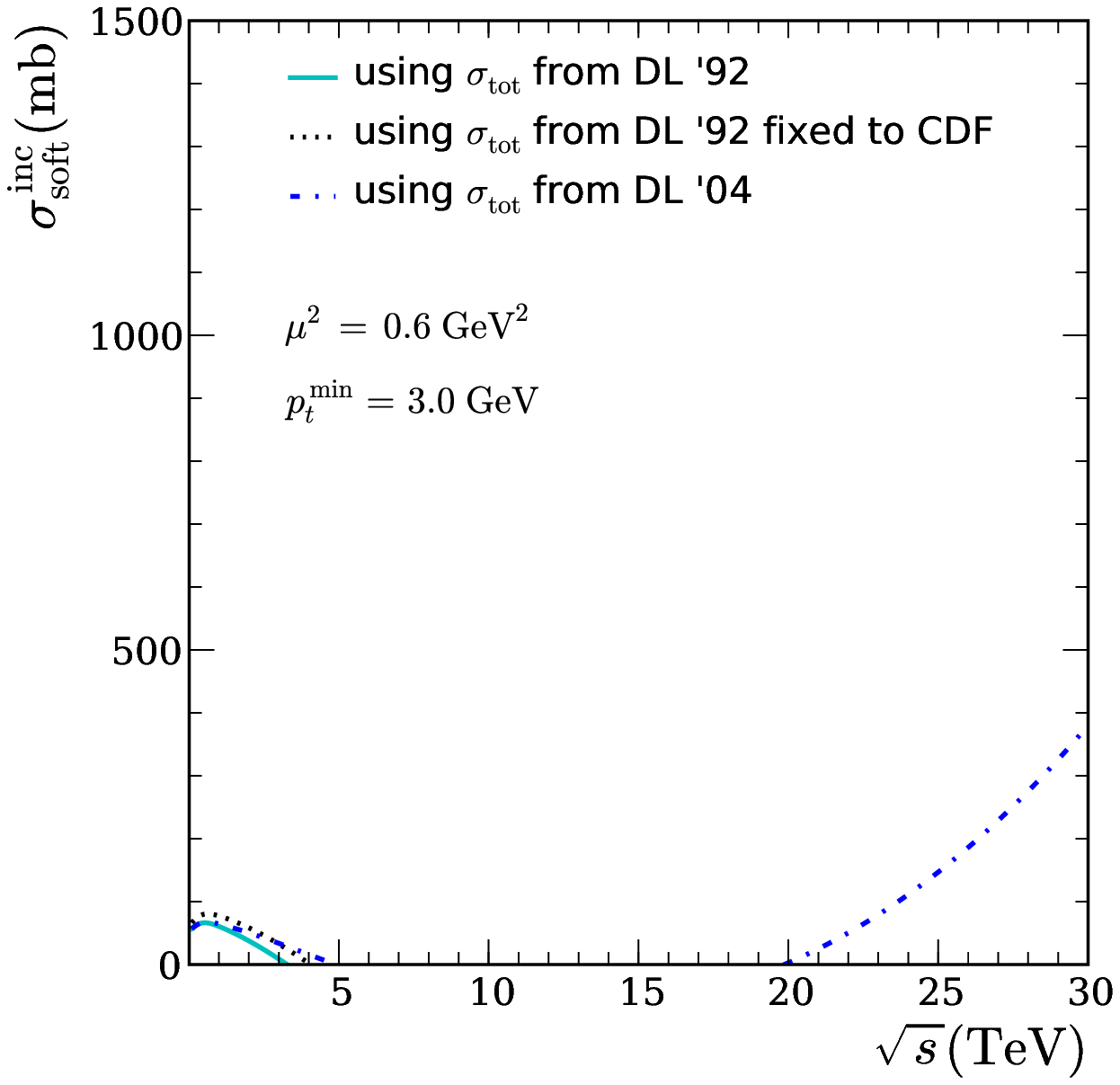}
    \includegraphics[width=0.47\textwidth]{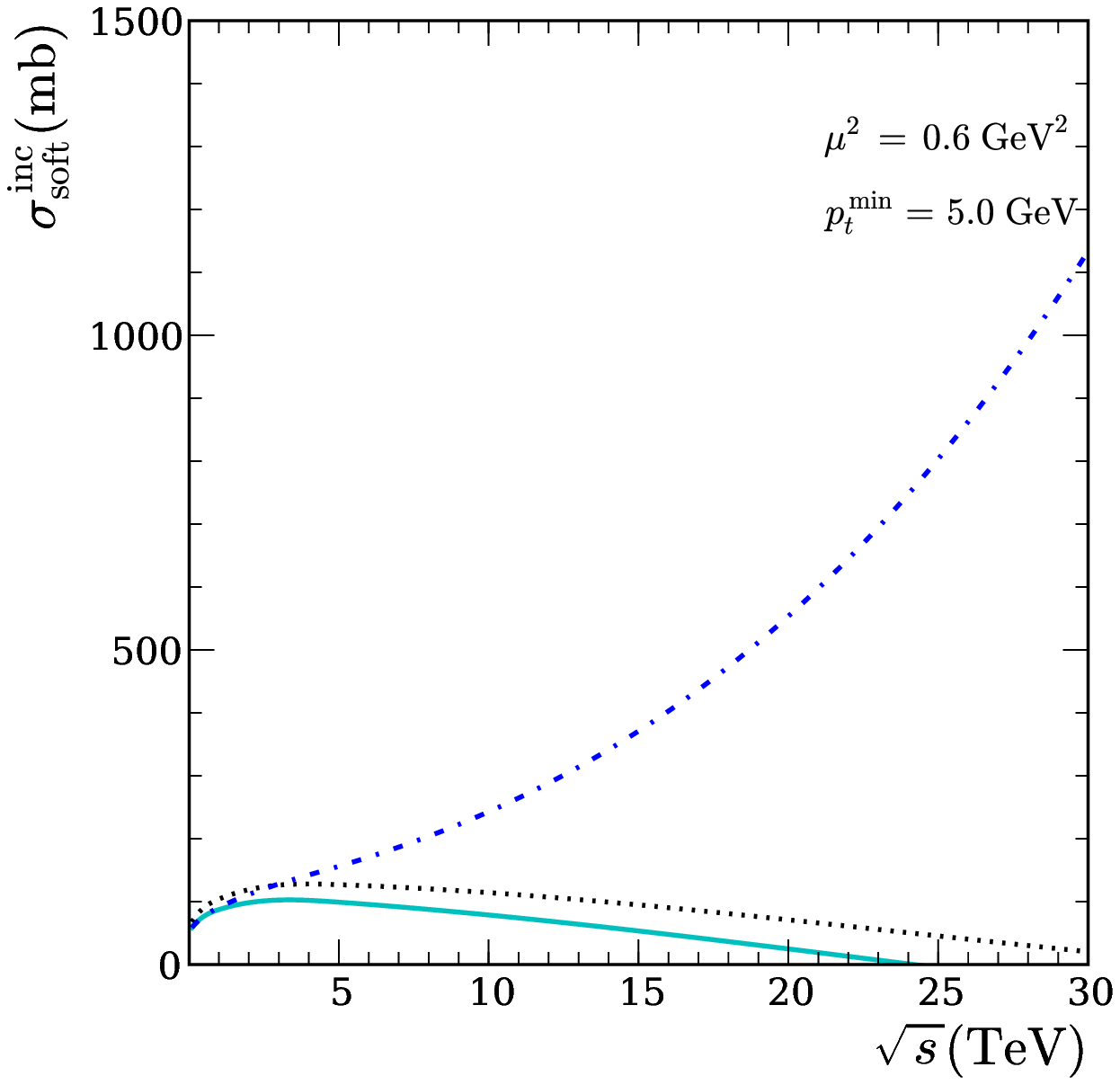}
  }\\
  \centerline{
    \includegraphics[width=0.47\textwidth]{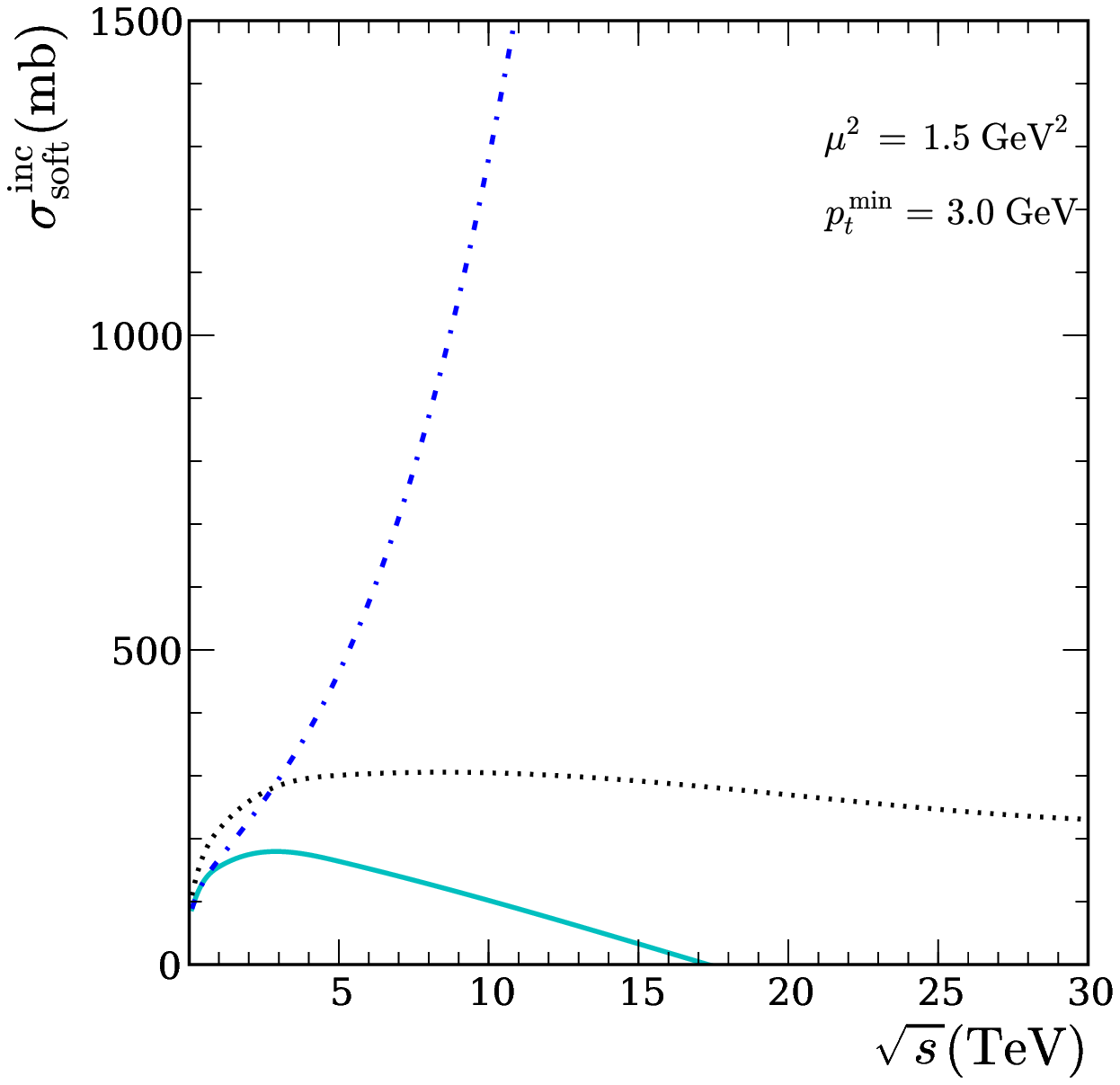}
    \includegraphics[width=0.47\textwidth]{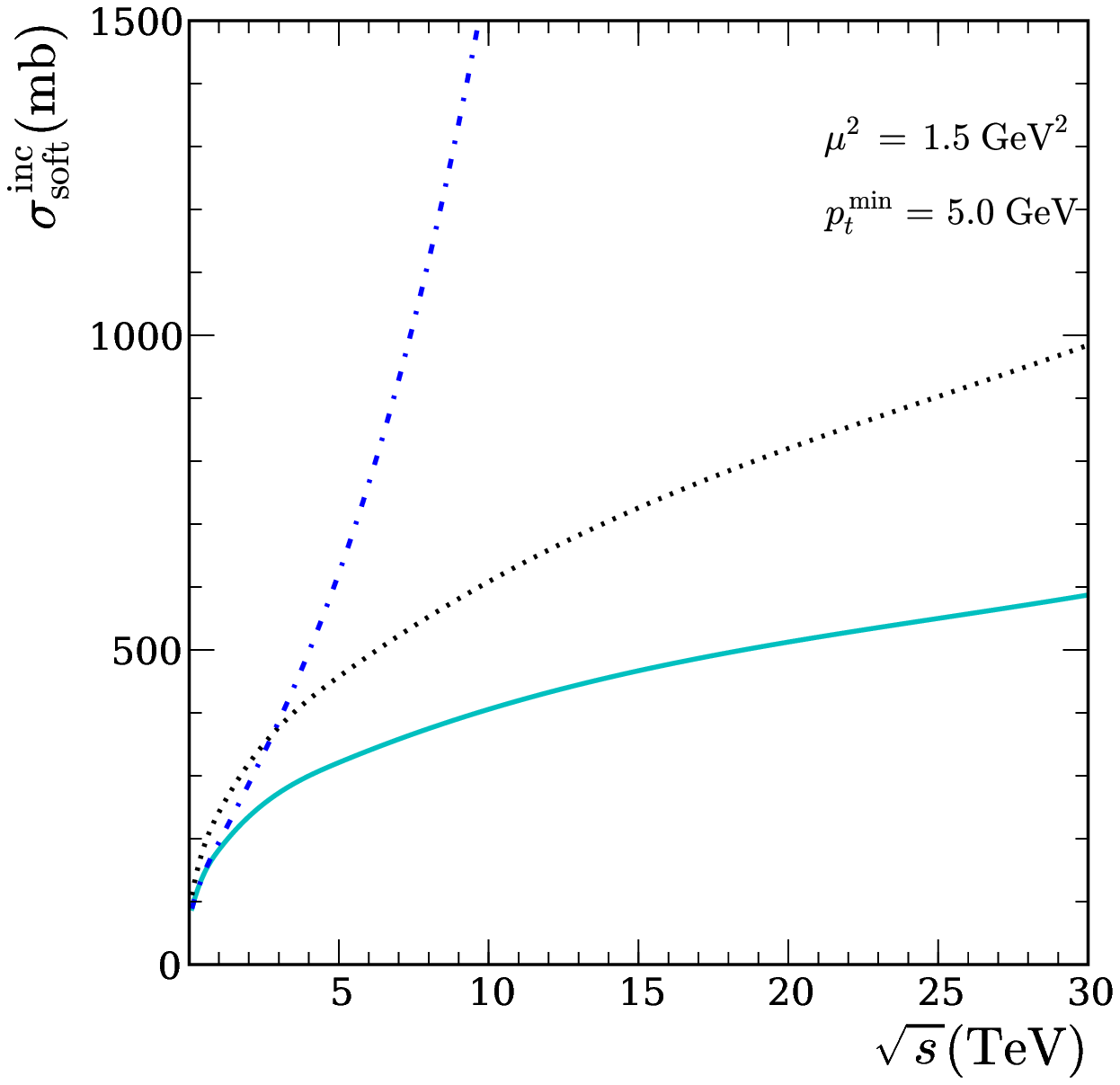}
  }
  \caption{$\sigmasoft$ for four different points in parameter
    space. As explained in the text, the extracted value of $\sigmasoft$
    depends on the values of $\sigmatot$, $\mu^2$ and (through the fact
    that $\sigmasoft=\sigma_{\rm de-eik}-\sigmahard$) $\ptmin$.  Each
    panel shows a different pair of $\mu^2$ and $\ptmin$ parameters,
    while the three different curves in each use the three
    parameterizations for the total cross section as a function of
    energy.}
  \label{fig:sigma_soft}
}

\subsection{Parton saturation physics}\label{sec:smallx}

The main motivation for allowing multiparton scatterings is the rise of the
inclusive cross section, for fixed $\ptmin$, with increasing
centre-of-mass energy. Multiparton scattering provides a mechanism to allow this 
quantity to exceed the total cross section. Eikonal models
that incorporate this fact unitarize the inclusive cross
section. There is however a second source of unitarization, the physics
of parton saturation, which is a competing effect. To estimate the
influence on our studies, we have used a simple modification of the PDFs
recently introduced \cite{Bahr:2008tx} into \HWPP\ to mimic parton
saturation effects for any PDF. The modification replaces $x f(x)$ below
$x_0$ by
\begin{align}
  x f(x) \to \left(\frac{x}{x_0}\right)^{\rm Exp} \ x_0 f(x_0)
  \quad \forall\; x < x_0 \, ,
\end{align}
where \HWPPParameter{SatPDF}{X0} and \HWPPParameter{SatPDF}{Exp} are
changeable parameters. HERA data indicate that saturation is unlikely
to be a strong effect above $x \approx 10^{-4}$. Therefore, the
strongest reasonable influence from this effect is obtained by setting
$x_0 = 10^{-4}, \rm{Exp} = 0$. The results are shown in Fig.~\ref{fig:deeik},
where the effect is visible, but small, at LHC energies.

\section{Parameter space constraints from data}

In discussing the de-eikonalized cross section, we noted that for some
parameter values the hard partonic cross section exceeds it. This implies
in our model that the soft cross section should be negative. We take
this as an inconsistency that would, for a given measured $\sigmatot$ at the LHC, 
rule out such parameter space points. In this section we 
discuss the extent to which the space of parameter values can be limited by this 
and other constraints.

\subsection{Consistency}\label{sec:Consistency}

The parameter space in $\mu^2$ and $\ptmin$ is shown in
Fig.~\ref{fig:constraintstev} for the Tevatron, and
Fig.~\ref{fig:constraintslhc} for the LHC.

The horizontal band shows the range of $\mu^2$ values allowed for a
given value of the elastic slope in conjunction with the indicated
$\sigmatot$. For Tevatron energies, both $B$ and $\sigmatot$ are chosen
according to the CDF measurement from Refs.~\cite{sigma_tot_CDF} and
\cite{Abe:1993xx} respectively.

Our expectations on the value of the elastic slope at LHC energies
simply reflect the range of predictions that the models of \cite{dl,
Donnachie:2004pi, Khoze:2000wk, Gotsman:2007ac} give, as discussed in
Sect.~\ref{sec:DL}. For the value of $\sigmatot$ at the LHC we show a
range of possible values motivated by these parameterizations.

For a particular value of $\sigmatot$ (or for a given range of possible
values at the LHC), we are able to extract constraints on the allowed
parameters, by simply requiring a sensible performance of the eikonal
model.  The most basic requirement, which was just mentioned,
is that the non-perturbative cross section that is needed to match the
total cross-section prediction is positive. A negative value means that
the model cannot be applied and therefore this requirement puts a
stringent limit on the allowed values of $\mu^2$ and $\ptmin$. This
limit will depend on the value of $\sigmahard$, which is not a stable
prediction itself. We therefore calculate this limit with several
variations. We use three different PDF sets \cite{Martin:2001es,
Pumplin:2002vw,Sherstnev:2007nd}, vary the running of $\alpha_s$ from
1-loop, which is the default in \HWPP\ to 2-loop and finally apply the
modifications to the PDF's described in Sect.~\ref{sec:smallx}.  The
solid lines in Figs.~\ref{fig:constraintstev} and
~\ref{fig:constraintslhc} show these limits, where the entire range
below the curves is excluded. The limits impose a minimal $\mu^2$ for
any given value of $\ptmin$. Points on that line are parameter sets
where $\sigmasoft = 0$~mb.

Another, weaker, consistency constraint we apply is related to the
simulation of the final state of these collisions.  We observe that when
we embed them into the full simulation of \HWPP, including backward
evolution of the initial state, each collision consumes, on average,
about a tenth of the available total energy, so that the approximation
that individual hard scatters are independent must break down, at least
due to energy conservation, when there are more than about ten of them.
We therefore indicate on Fig.~\ref{fig:constraintslhc} the points in
parameter space where the average multiplicity of scatters above
$\ptmin$ reaches 10.  This is certainly not a stringent limit but a
sensible parameter choice most likely avoids this region.

The classic CDF analysis of the distribution in azimuth of the mean
charged multiplicity and scalar $\pt$ sum as a function of the
transverse momentum of the leading jet\cite{Affolder:2001xt} also
provides constraints on the model as embedded in \HWPP.
Reference~\cite{Bahr:2008dy} described the implementation of multiparton
scattering into \HWPP\ (i.e.~the simulation of the final state
corresponding to $\sigmahard$) and made a two-parameter fit ($\mu^2$ and
$\ptmin$) to these data.  Since \HWPP\ does not yet include a simulation
of the final state corresponding to $\sigmasoft$, we do not take the
results of this fit as a strong constraint on the parameter space, but
an indication of the effect such a tuning could have once a complete
description is available.  The result is that, although one obtains a
best fit with the values $\mu^2=1.5\GeV^2$, $\ptmin=3.4\GeV$, the
best-fit values of the parameters are strongly correlated, with the
$\chi^2$ function having a long, thin, rather flat valley running from
$(\ptmin=2.5\GeV, \mu^2\sim0.7\GeV^2)$ to $(\ptmin=4.5\GeV,
\mu^2\sim2.5\GeV^2)$, and beyond. For any given value of $\ptmin$ in
this range one can find a $\mu^2$ value that gives a good description of
these data.

\FIGURE[t]{
  \vspace*{-0.2cm}
  \includegraphics[width=0.48\textwidth]{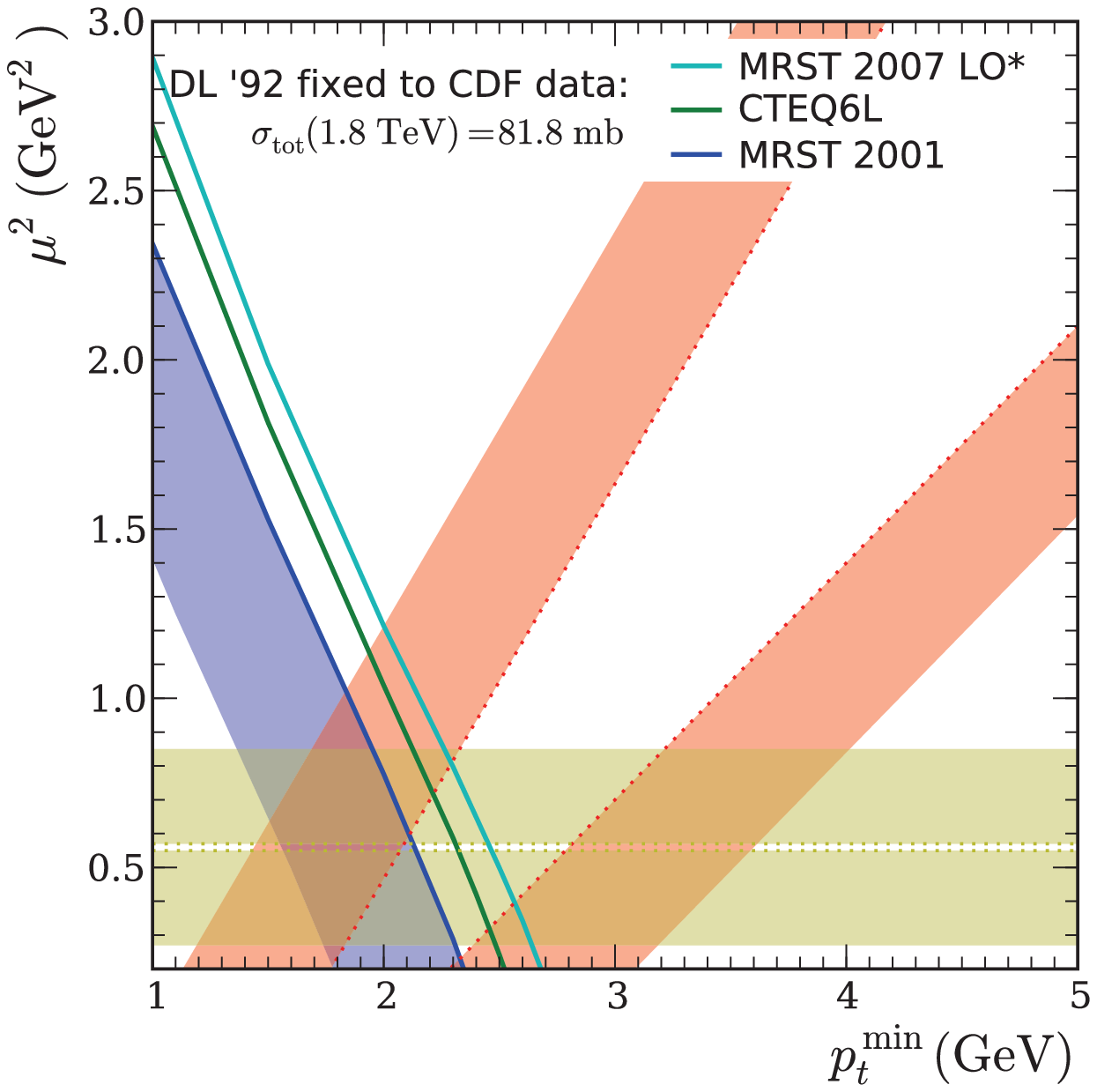}
  \includegraphics[width=0.48\textwidth]{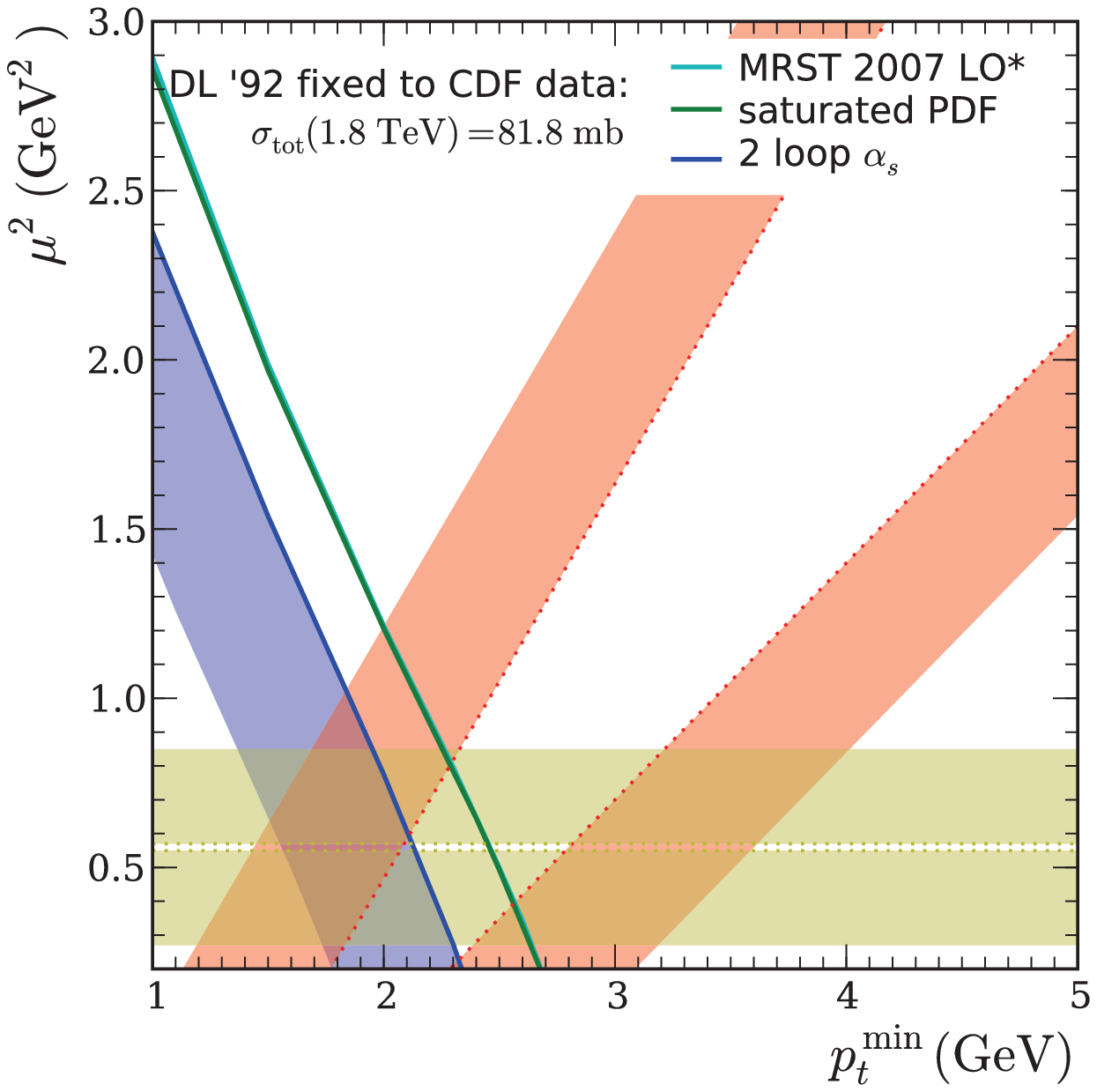}
  \vspace*{-0.2cm}
  \caption{The parameter space of the eikonal model at
    Tevatron energies. The solid curve imposes a minimum allowed value
    of $\mu^2$, for a given value of $\ptmin$ by requiring a positive
    value of $\sigmasoft$. The horizontal lines correspond to the
    measurement of $B = 16.98 \pm 0.25 \GeV^{-2}$ from CDF
    \cite{sigma_tot_CDF}. The excluded regions are shaded. The dashed
    lines indicate the region of preferred parameter values for a
    fit to Tevatron final-state data from Ref.~\cite{Bahr:2008dy}, which
    used the MRST2001 PDF set. The left plot shows the PDF uncertainty
    by varying the PDF set. The right plot shows the uncertainty that is
    implied by using 2-loop $\alpha_s$ running and using the saturation
    modifications.}
  \label{fig:constraintstev}
} 

Combining these constraints at the Tevatron, a small allowed region
remains around $\ptmin = 2.3 \GeV$ and $\mu^2 = 0.6 \GeV^2$.

At the LHC, this region would be ruled out for all the values of
$\sigmatot$ we have considered.  Note that if the LHC measurement were
as high a 164~mb, this would on its own imply an energy-dependent
$\mu^2$, in contradiction with our initial assumptions.

In the next section we discuss different ways in which the assumptions
of the model might be modified to account for this potential
inconsistency.

\FIGURE[htb]{
  \vspace*{-0.5cm}
  \centerline{
    \includegraphics[height=0.28\textheight]{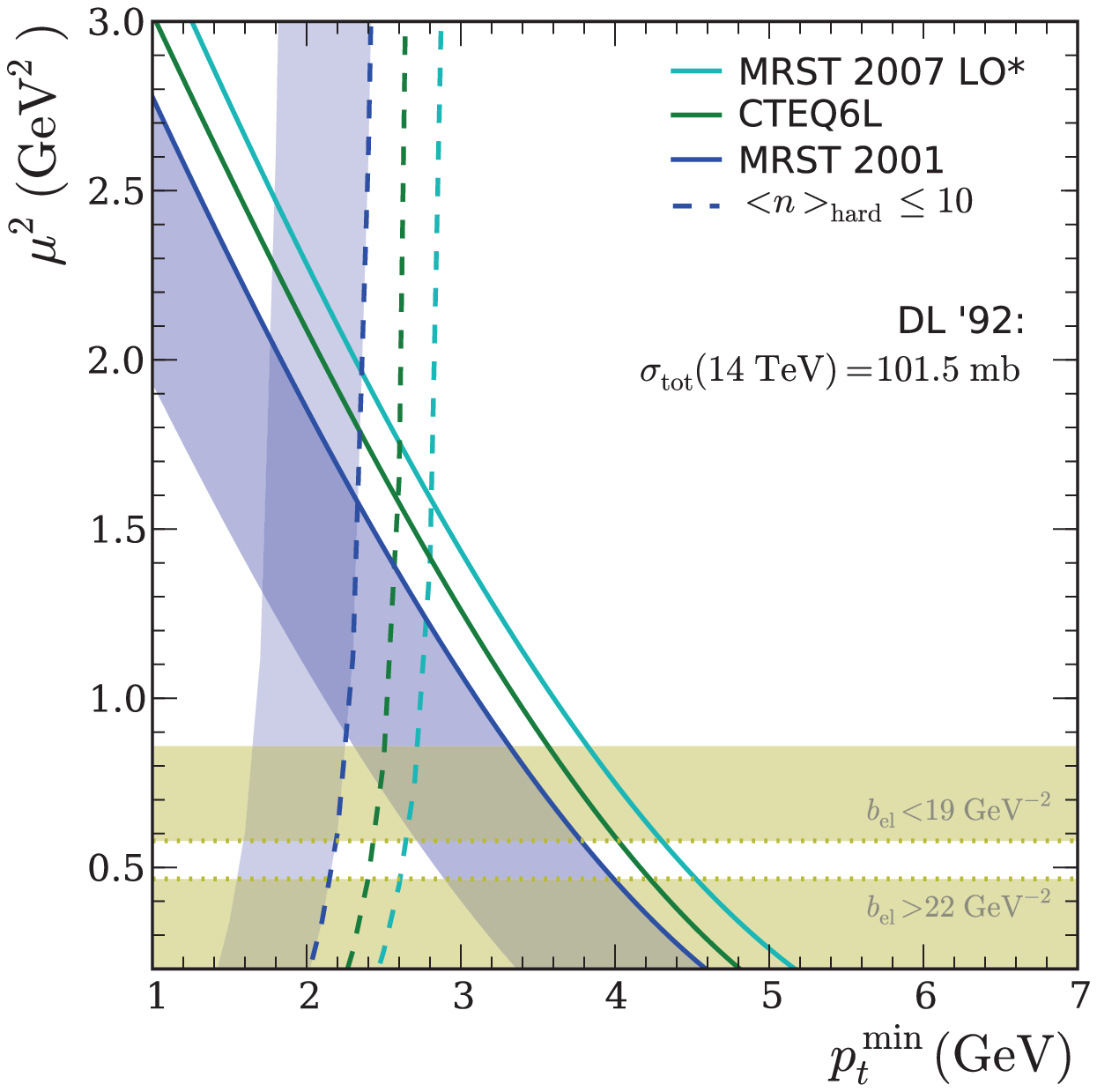}
    \includegraphics[height=0.28\textheight]{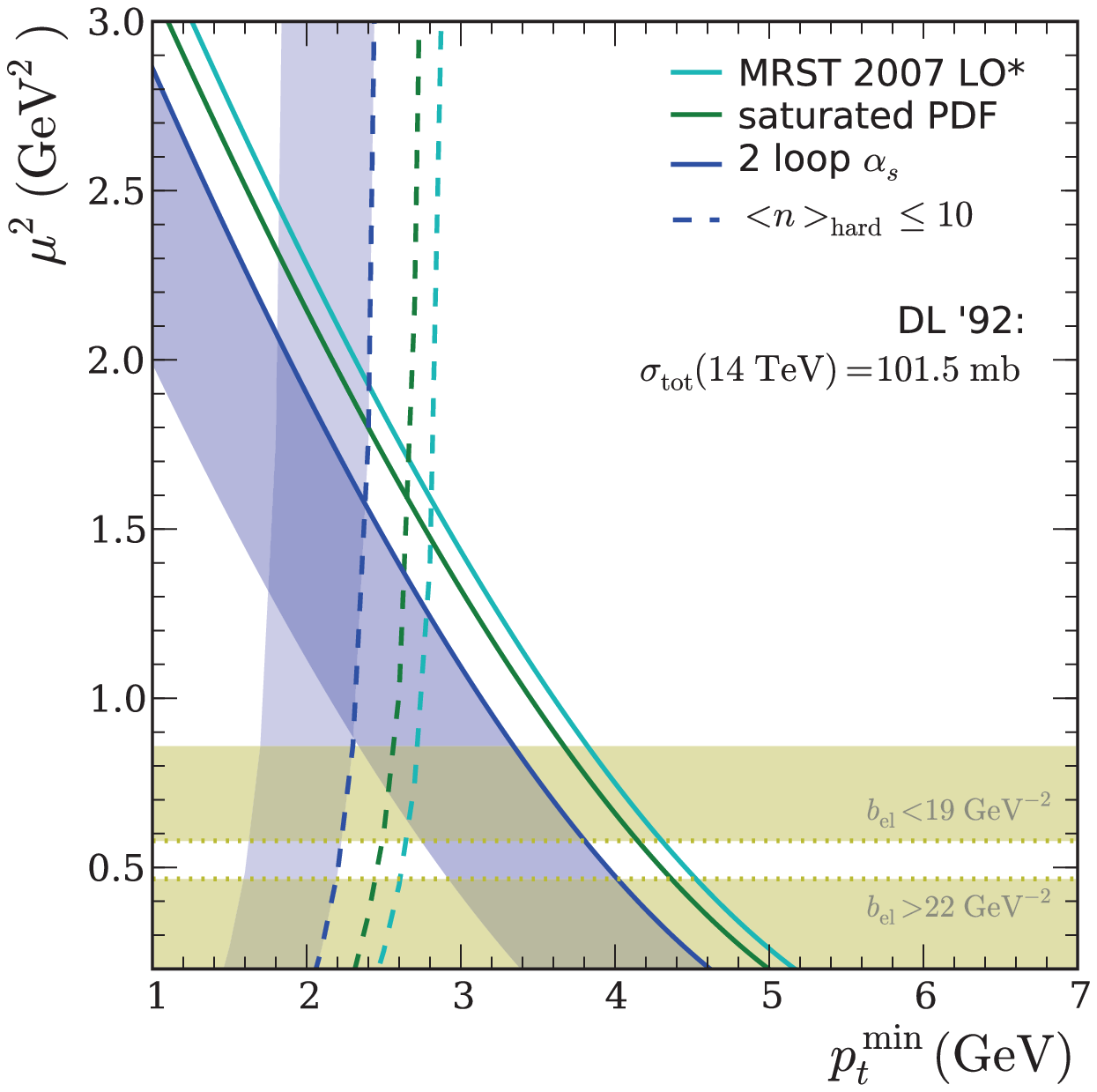}
  }\\
  \centerline{
    \includegraphics[height=0.28\textheight]{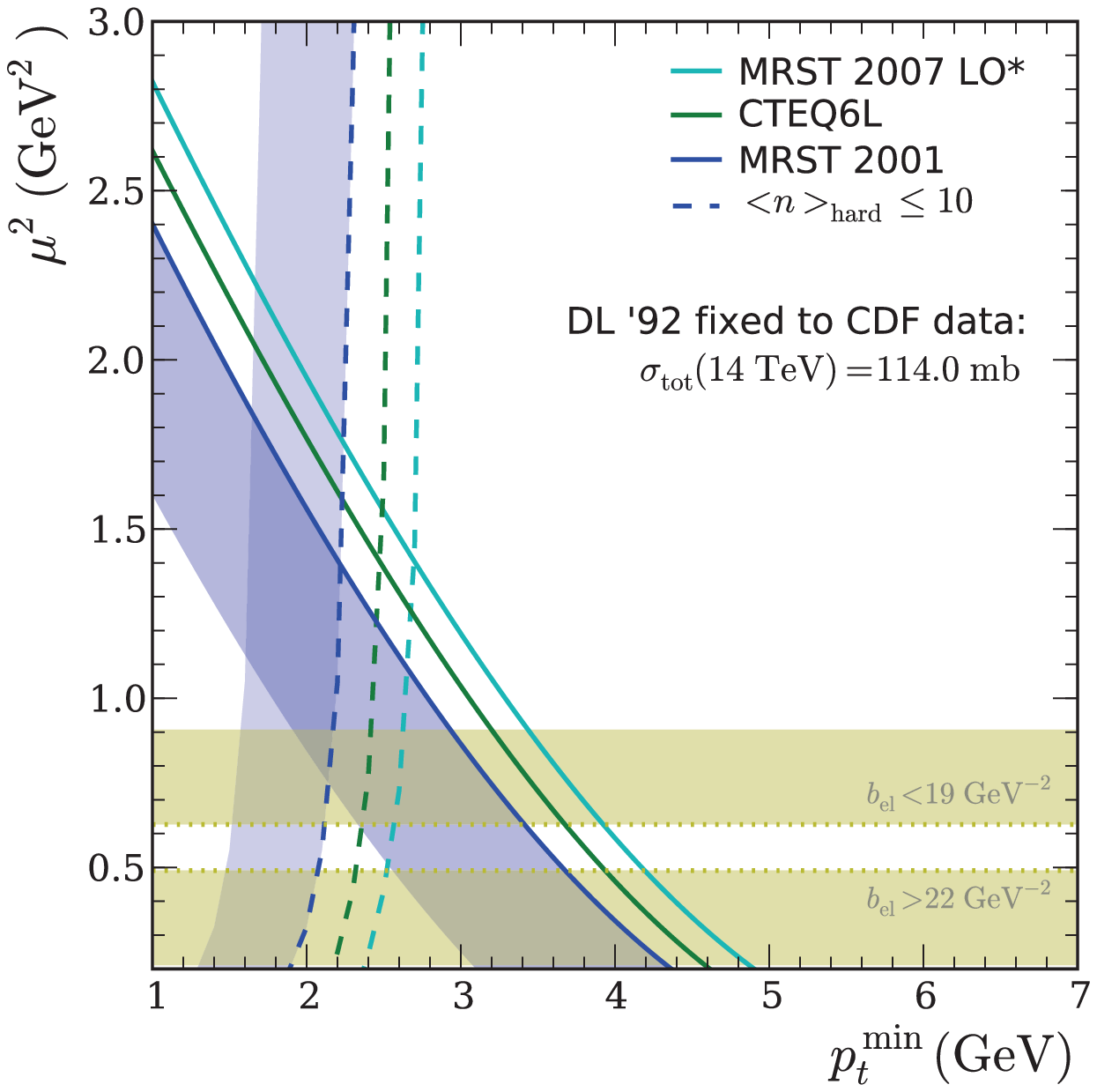}
    \includegraphics[height=0.28\textheight]{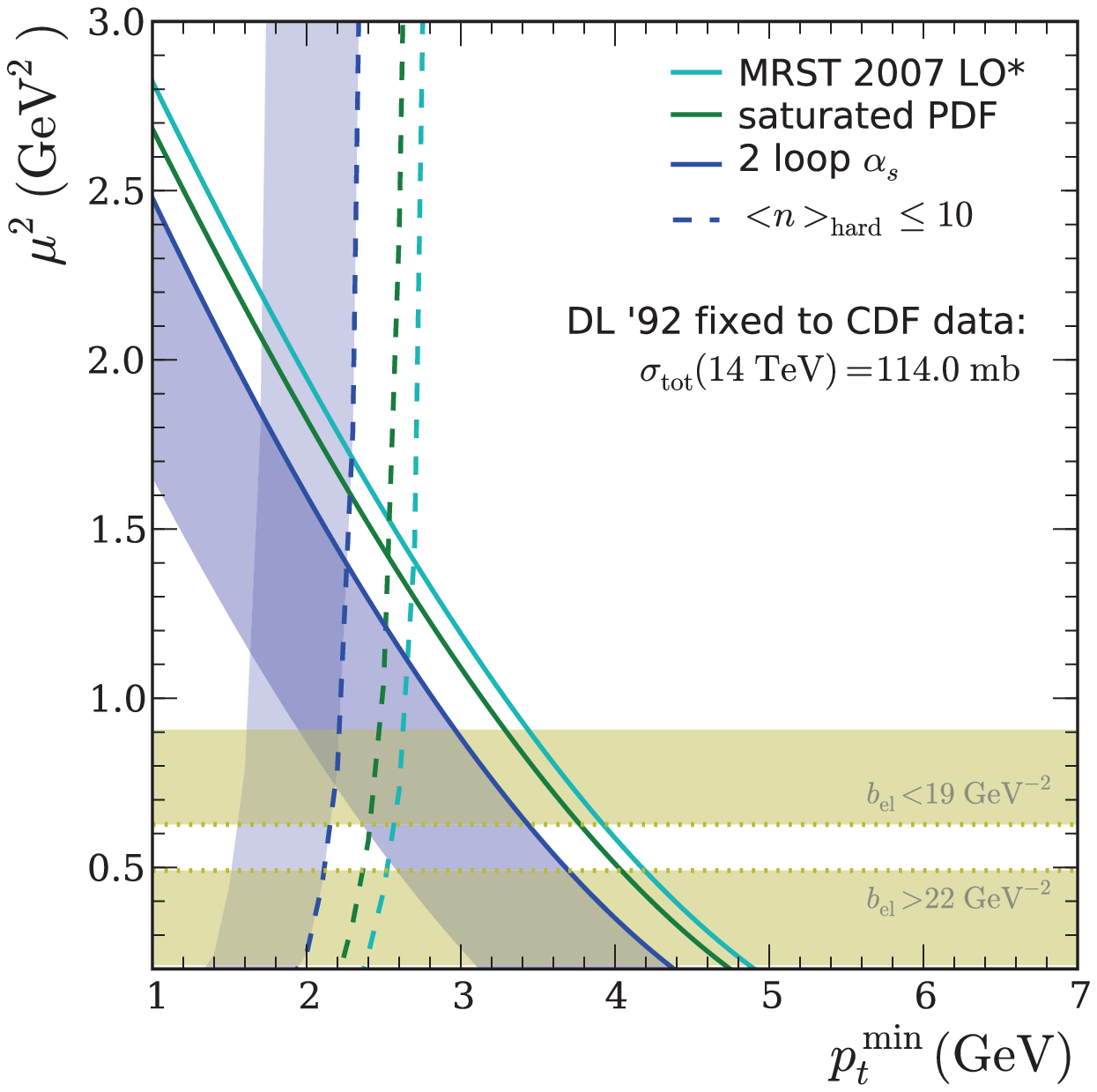}
  }\\
  \centerline{  
    \includegraphics[height=0.28\textheight]{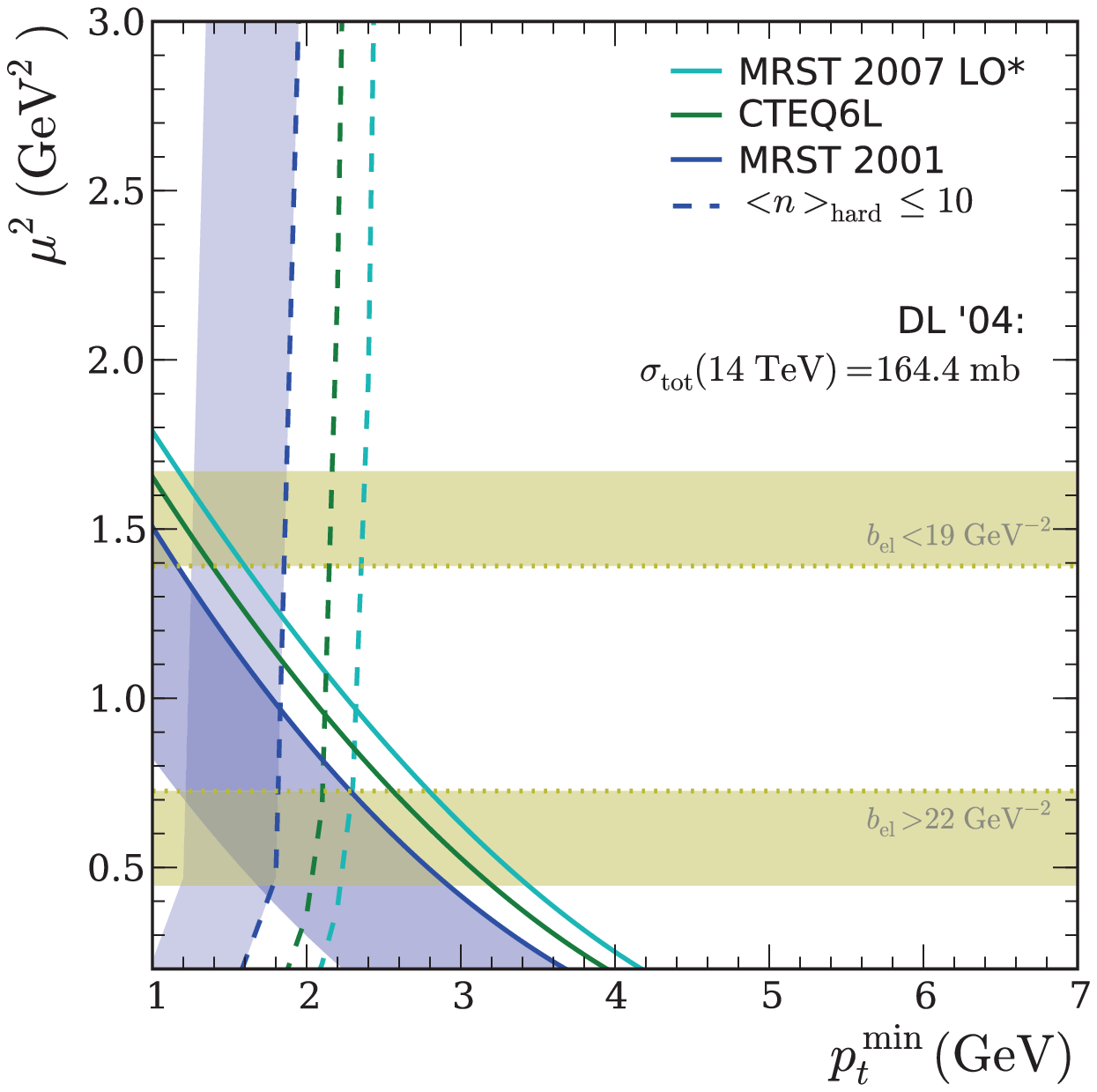}
    \includegraphics[height=0.28\textheight]{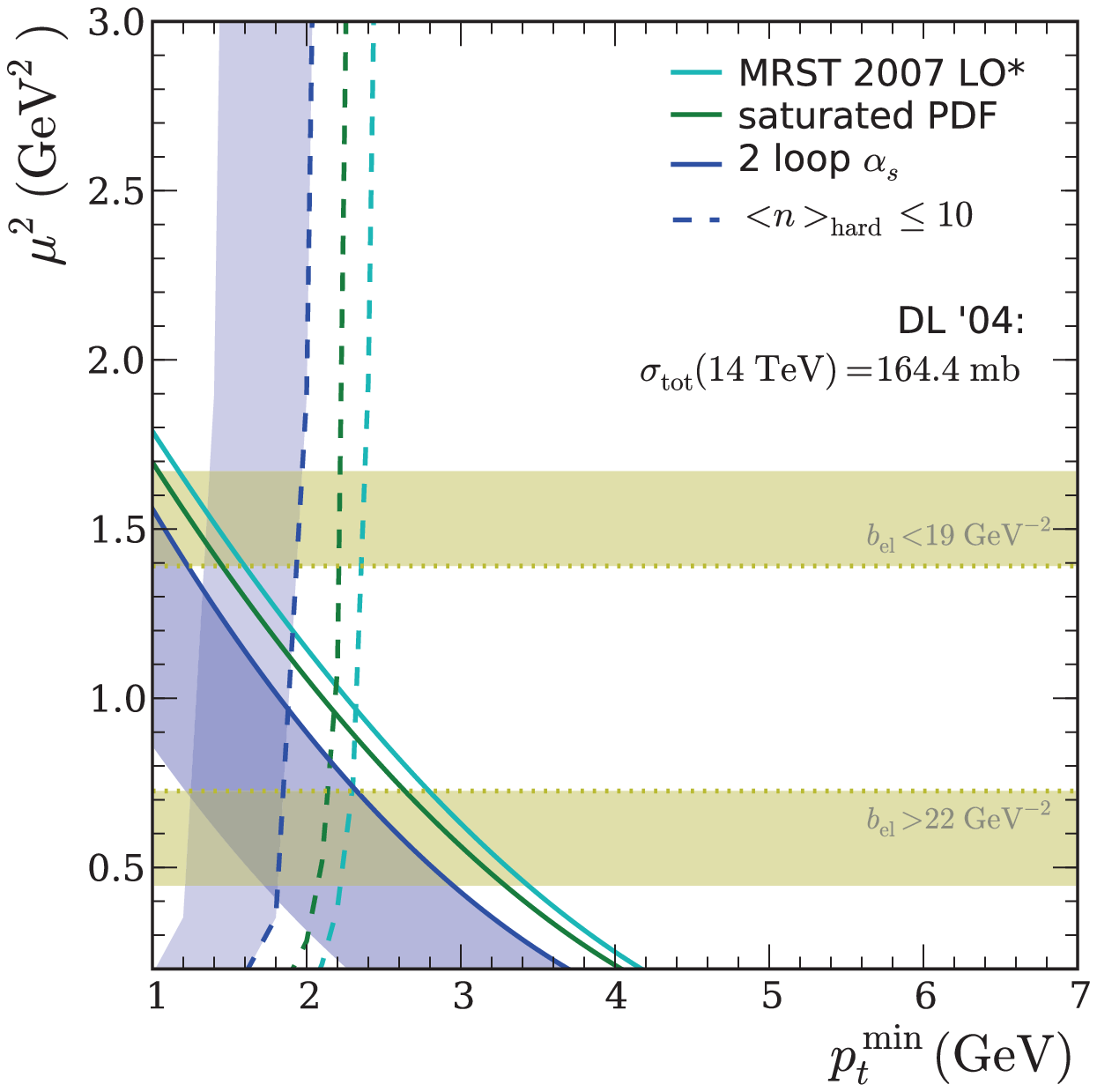}
  }
  \vspace{-0.5cm}
  \caption{The parameter space of the eikonal model and
    three constraints. The first one drawn as solid curve imposes a
    minimum allowed value of $\mu^2$, for a given value of $\ptmin$ by
    requiring a positive value of $\sigmasoft$. The second one, in
    dashed lines is deduced from an upper limit of the average number of
    additional semi-hard scatters in a typical minimum bias event. The
    excluded regions are shaded. The third constraint comes from the
    expected range of values for the elastic slope $B$. From top to
    bottom we calculate these constraints for the three different total
    cross sections at LHC, discussed in Sect.~\ref{sec:DL}, always with
    the same range of $B=19-22\GeV^{-2}$. Finally the
    left column shows the PDF uncertainty by varying the PDF set. The
    right column shows the uncertainty that is implied by using 2-loop
    $\alpha_s$ running and using the saturation modifications.}
  \label{fig:constraintslhc}
} 

\subsection{Extensions to the model}\label{sec:energydep}

Some authors have suggested, within multiparton scattering models, that
the parameters of the model, analogous to our $\mu^2$ and $\ptmin$,
should be energy dependent. In this section we briefly discuss the
arguments for these models.  

In \cite{Dischler:2000pk} a simple model of the spatial/momentum
structure of a hadron was constructed from which it was argued that the
colour screening length decreases slowly with increasing energy.  This
translates into a $\ptmin$ that increases slowly with energy, for which
they estimated $\ptmin\sim s^\epsilon$ with $\epsilon$ in the range
0.025 to~0.08.  The actual value used in
Refs.~\cite{Sjostrand:2004pf,Sjostrand:2006za,Sjostrand:2007gs} is 0.08,
leading to a 35\% increase in $\ptmin$ from the Tevatron to the LHC.

In \cite{Godbole:2004kx,Achilli:2007pn} a multiparton model was
constructed that is very similar to ours at low energy, with an impact
parameter distribution of partons given by the electromagnetic form
factor.  However, soft gluon effects were estimated and summed to all
orders, to give a mean parton-parton separation, $b_{rms}$, that falls
with energy, quickly at first, but then saturating: the value at 1~TeV
is about a factor of two smaller than at low energy, while the value at
14~TeV is only about 10\% smaller still.  In terms of our simple model
in which the matter distribution always has the form factor form and is parameterized by
$\mu^2$, $\langle b^2\rangle \propto 1/\mu^2$ and this corresponds to $\mu^2\sim2.8\GeV^2$ 
at the Tevatron and $\sim3.4\GeV^2$ at the LHC. Not only would this introduce an energy 
dependence in $\mu^2$, but the values imply a different $\mu^2$ for hard partonic interactions 
than that derived from the measured elastic slope parameter, a point
that we will return to below.

Note that both these sources of energy dependence would
act in the right direction to evade the potential consistency
constraints at the LHC.  Allowing
$\ptmin$ and/or $\mu^2$ to increase with energy would move the model towards the allowed region in 
Fig.~\ref{fig:constraintslhc}.

The CDF collaboration have published measurements of the double-parton
scattering cross section\cite{Abe:1997bp,Abe:1997xk}.  As pointed out in
Ref.~\cite{Treleani:2007gi} the quantity called $\sigma_{\rm eff}$
there is not the effective cross section as it is usually defined,
\begin{equation}
  \sigma_{\rm eff} = \frac1{\int d^2\vect{b}
  \bigl(A(b)\bigr)^2},
\end{equation}
but is related to the latter by a small correction.  Using the value of
this correction estimated in Ref.~\cite{comment}, we obtain
$\sigma_{\rm eff}\sim11.5\pm2$~mb.  In our form
factor model, this corresponds to $\mu^2\sim3.0\pm0.5\GeV^2$.  It is
interesting to note that this value is close to the one predicted by the
analysis of Refs.~\cite{Godbole:2004kx,Achilli:2007pn} mentioned earlier.
Again, this value is inconsistent with the assumption that the hard scatters 
``see'' a form factor matter distribution derived from the elastic slope parameter.
Recall from our earlier discussion that we do not expect significant qualitative differences
for other models of the matter distribution in the proton, merely some
distortions of the parameter-space plane. 

Improved analyses of these and other observables are under way and, once
completed, in particular with a simulation of the final state of
$\sigmasoft$, will provide strong constraints on the values of the
parameters $\mu^2$ and $\ptmin$ in our model. 

\section{Conclusions}

The connections between our underlying event model and the total
proton-proton cross-section have been discussed.  Requiring consistency
of the model up to LHC energies imposes constraints on the allowed
parameter values, for a given range of possible measurements of
$\sigmatot$ at the LHC.  Our main result is summarized in
Fig.~\ref{fig:constraintslhc}, which shows these constraints for various
values of the total cross section at the LHC and various inputs to the
perturbative cross section calculation. Taking the Tevatron data
together with the wide range of possible values of $\sigmatot$
considered at LHC, no allowed set of parameters ($\mu^2$, $\ptmin$)
remains for our simple model.

This would imply that soft and hard scatters see different matter
distributions as a function of impact parameter and/or that the
parameters of our model are energy dependent. The measurement of double-parton
scattering at the Tevatron supports the idea that hard scatters see a
more dense matter distribution than is implied by the $t$-slope of the
elastic cross section. Various phenomenological models also predict such
effects.

With steadily improving data from the Tevatron, more detailed
phenomenological analyses being performed and the prospect of data from
the LHC, there is a real hope that the various simplifying assumptions
that go into our model will be tested to the limit and we will discover
where, if anywhere, more detailed understanding of the dynamics of
underlying event physics is needed. The biggest unknown in our analysis
is the total cross section at the LHC. With even a first imprecise
measurement of this cross section, we could strengthen our parameter
space analysis considerably.

\section*{Acknowledgements}

We are grateful for discussions with Stefan Gieseke and Jeff Forshaw.
This work was supported in part by the European Union Marie Curie
Research Training Network MCnet under contract MRTN-CT-2006-035606. MB
was supported by the Landesgraduiertenf\"orderung Baden-W\"urttemberg.

\bibliographystyle{JHEP}
\bibliography{literature}

\end{document}